\documentclass[a4paper,11pt]{article}
\usepackage{pos}
\usepackage{setspace}

\usepackage{comment}
\usepackage{cleveref}
\newcommand\lf[1]{{#1}}

\usepackage{graphicx}
\usepackage[caption=false]{subfig}	

\title{Neutrinos from interactions between the relativistic jet and large-scale structures of BL~Lac objects investigated through their gamma-ray spectrum}
 \ShortTitle{Neutrinos from interactions between the BL~Lac jet and its large-scale structures}

\author*[a]{L. Foffano}
\author[b]{M. Cerruti}
\author[a]{V. Vittorini}

\affiliation[a]{INAF - IAPS, via del Fosso del Cavaliere 100, I-00133 Roma (Italy)}

\affiliation[b]{Université Paris Cité, Astroparticule et Cosmologie (APC) 10, rue Alice Domon et Léonie Duquet
75013, Paris (France)}

\emailAdd{luca.foffano@inaf.it}

\abstract{Absorption and emission lines in the optical spectrum are typically used to investigate the presence of large-scale environments in active galactic nuclei (AGNs). BL~Lac objects - which are a category of AGNs with the relativistic jet pointing directly to the observer - are supposed to represent a late evolution stage of AGNs. Their large-scale structures are probably poorer of material, which is distributed with lower densities throughout the circumnuclear environment. Their accretion disk is weak and weakly reprocessed, making the non-thermal continuum of the relativistic jet dominate their optical spectrum and preventing us from identifying the thermal emission of the photon fields produced by such large-scale structures. However, these photon fields may still exist and eventually interact with the gamma rays traveling in the blazar jet via gamma-gamma pair production, producing observable effects such as absorption features in their spectral energy distribution.
Interestingly, the same photon field might also lead to the production of high-energy neutrinos, acting as targets for proton-photon interactions. In this contribution, we present the results of a set of simulations over a wide parameter space describing both the blazar jet and the photon field properties. We discuss the most effective conditions that may produce fluxes of neutrinos compatible with the sensitivities of the current and the next generation of neutrino detectors. We will also discuss how the possible neutrino flux would be related to the properties of the large-scale structures investigated indirectly through the analysis of the gamma-ray spectrum of the BL~Lac object. 
}

\ConferenceLogo{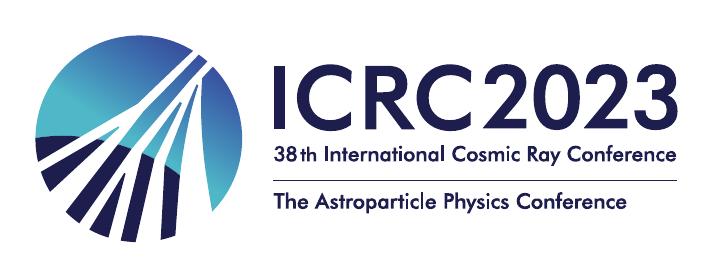}

\FullConference{%
38th International Cosmic Ray Conference (ICRC2023)\\  26 July - 3 August, 2023\\  Nagoya, Japan}

\begin{document}
\maketitle

\section{Introduction}
\noindent
Blazars constitute a category of active galactic nuclei (AGNs) emitting relativistic jets directed toward the line-of-sight of the observer. Among them, BL~Lac objects are blazars in which the emission of thermal components is overshadowed by the non-thermal radiation of the jet. This jet radiates across the entire electromagnetic spectrum, extending into the gamma-ray energy range. The existence of large-scale structures within the AGN environment is commonly explored through the presence of ambient photon fields emitting in its optical spectrum. However, in BL~Lac objects, this thermal emission is also commonly hidden by the non-thermal continuum of the relativistic jet, making it challenging to identify large-scale structures within these objects.

\section{Absorption features in gamma-ray spectra of BL Lac objects}
\noindent
An alternative method for investigating the potential presence of large-scale structures around BL~Lac environments has been proposed by \citet{Foffano2022}. In that work, we develop an alternative technique aimed at unveiling large-scale structures through the analysis of the gamma-ray spectrum of the blazar.
We assume that the relativistic jet of a BL~Lac object is passing through a dense photon field produced by the large-scale structures of the blazar. \lf{To simplify the discussion, we identify the representative large-scale structure with a narrow-line region (NLR). However, it is crucial to emphasize that the method is based on the total photon column density provided by the target photon field. Consequently, it can be applied to any configuration of clouds within the medium- and large-scale structures of the AGN that can supply the necessary photon column density.}

If gamma rays are produced before such a dense target photon field, their observed flux will be severely affected (or even completely suppressed) by the $\gamma\gamma$ interaction $\gamma_{\text{jet}} + \gamma_{\text{target}} \to e^+ + e^-$ \citep[for more details see][]{aharonian2004book}, following 
\begin{equation*}
I_{\text{out}}  = I_{\text{in}} \: e^{-\tau_{\gamma\gamma}} \; ,
\label{eq:absorption-flux}
\end{equation*}
where $I_{\text{out}}$ and $I_{\text{in}}$ represent the observed and the original flux of gamma rays, respectively.  The absorption factor $\tau_{\gamma\gamma}$ of the $\gamma\gamma$ interaction  - assuming a mono-energetic, isotropic, and uniform seed photon field - can be expressed as a function of the cross-section $\sigma_{\gamma\gamma}(E)$, of the size of the interacting region $R$, and of its photon density $n_{\text{seed}}$: $\tau_{\gamma\gamma} = \; n_{\text{seed}} \cdot R \cdot \sigma_{\gamma\gamma}(E)\label{eq:tau-definition}$.
Specifically, we define the photon column density $K_{\text{seed}} = n_{{\text{seed}}} \cdot R$, which is a critical parameter describing the target photon field and that is related to the strength of the absorption process.
When dealing with conventional AGN photon fields with energies spanning from infrared to optical and UV, the $\gamma\gamma$ interactions produce the strongest absorption at gamma rays, making this process indirectly detectable by our gamma-ray detectors. \lf{The observable effect is that in the gamma-ray spectrum of candidate sources, the typical inverse Compton emission is influenced by an absorption feature. The typical energy of this feature relies on the energies of the interacting photons and its maximum intensity depends on the photon column density of the target photon field.}

This process is being investigated with a systematic analysis of the gamma-ray spectra of numerous BL~Lac objects.
Interestingly, the most promising sources where this effect could be easily detectable are high-peaked BL~Lac objects (HBLs) and extreme HBLs (EHBLs, or \emph{extreme blazars}, e.g. \citet{foffano-2019, costamante_2001}), which are BL~Lac objects defined on the basis of their synchrotron peak frequency lying between $10^{15}$ and $10^{17}$ Hz for HBLs and above $10^{17}$~Hz for EHBLs. Thanks to their strong and uncontaminated emission up to TeV energies, they offer the best spectra to investigate the presence of gamma-ray absorption features.

\section{Neutrino production in BL~Lac objects}

\noindent
In this work, we explore the scenario proposed by \citet{Foffano2022}, 
examining the photo-meson interactions involving a distribution of protons accelerated in the jet and the local photon field produced by the large-scale structure. Such a photon field is also responsible for the absorption feature in the gamma-ray spectrum discussed above, and could serve as a target for photo-meson interaction eventually producing high-energy neutrinos.

The photo-meson (or photo-pion) process is the production of pions in proton-photon interactions \citep{Cerruti:2020lfj}:
\[
p + \gamma \to p_0 + \pi_0 \; \quad\;
p + \gamma \to n + \pi^+ \;  \quad\;
p + \gamma \to p_0 + \pi^+ + \pi^-  \; .
\]
For astrophysical applications, we mostly consider the production of neutral and charged pions, which can happen when the photon energy in the proton rest frame exceeds $\sim$145 MeV. 
Pions  decay into leptons (muons and electrons/positrons) and neutrinos following 
\[
\pi_0 \to \gamma\gamma  \; \quad\;   {\pi}^{\pm} \rightarrow {\mu}^{\pm}+{\nu}_{\mu}({\bar{\nu}}_{\mu}) \rightarrow e^{\pm}+{\nu}_{e}({\bar{\nu}}_{e})+{\nu}_{\mu}+{\bar{\nu}}_{\mu} \; .
\]

\noindent
A crucial aspect of photo-meson reactions is that the product is composed by neutrinos and photons together. Interestingly, neutrinos can escape from the emitting region without being affected by absorption effects. Then, detecting neutrinos and photons together from an AGN may be a smoking-gun signal for the presence of relativistic protons accelerated in the jet, making AGNs possible accelerators of protons to high energies and consequently a possible source of cosmic rays detected at Earth \citep[e.g.][]{2018arXiv180708794I, 2019MNRAS.489.4347O}.

\noindent
We are currently studying two main approaches:
\begin{description}
\item[Optical emission.]
In recent times, detailed monitoring of optical spectra of BL Lac objects also during periods of enhanced activity has shown that occasionally optical emission and absorption lines appear \citep[e.g.][among other works]{masquerading_bllac}. Then, assuming that such an optical emission is related to the photon field produced by the large-scale structure, we use it to compute neutrino flux and spectrum.
\item[Direct connection with the gamma-ray absorption feature.]
The stronger the $\gamma\gamma$ interaction between high-energy photons of the jet and low-energy photons of the target photon field, the more intense the absorption feature in the spectrum of the BL Lac object becomes, and the higher the probability that photo-meson reaction may occur and produce neutrinos. This assumption is based on the same connection between the absorption factor given by the $\gamma\gamma$ interaction and the efficiency of the $p\gamma$ reactions discussed in literature with different applications \citep[e.g.][and references therein]{2016PhRvL.116g1101M}.
\end{description}

\noindent
In both cases, we are computing the neutrino production rate and the neutrino spectrum \citep[following e.g.][]{2006PhRvD..73f3002M} arising from the interaction between the relativistic jet of the BL~Lac object and the possible local photon fields produced by its large-scale structures. Parameters of the model are related to the physical conditions of the relativistic jet, to the proton distribution, and to the properties of the target photon fields. 

To comprehensively address the subject, we are conducting a series of simulations that span the entire parameter space.
Concerning the relativistic jet, the primary parameters describe the usual properties of the blazar emitting region, such as the bulk Lorentz factor $\Gamma$, the magnetic field $B$, the radius $r$, and the redshift of the source $z$. 
Conversely, the properties of the photon field are based on the typical  spectral emission of large-scale structures investigated also in other categories of AGNs, for example radio galaxies.

Our goal is to explore the correlation between the potential neutrino flux and the properties of large-scale structures of BL~Lac objects, whose presence is indirectly investigated through the analysis of the gamma-ray spectrum of the blazar. Ultimately, we aim to determine the optimal conditions that could generate neutrino fluxes compatible with the sensitivities of both the current and next generation of neutrino detectors.
 
\qquad\\

\noindent
\textbf{Acknowledgements.} Research carried out through partial support of the ASI/INAF AGILE contract I/028/12/05 and of the European Project ``3rd AHEAD2020 Visitors Programme call''.

\setstretch{0.5}

\bibliography{bib_icrc2023}
\bibliographystyle{aasjournal}

\end{document}